\documentclass[review]{elsarticle}

\usepackage{lineno,hyperref}
\modulolinenumbers[5]

\journal{Information Sciences}

\usepackage{changepage}   
\usepackage{sidecap}
\usepackage{frcursive}
\usepackage{setspace}
\usepackage{caption}
\usepackage{epstopdf}

\usepackage{titlesec}

\titleclass{\subsubsubsection}{straight}[\subsection]

\newcounter{subsubsubsection}[subsubsection]
\renewcommand\thesubsubsubsection{\thesubsubsection.\arabic{subsubsubsection}}

\titleformat{\subsubsubsection}
  {\normalfont\normalsize\bfseries}{\thesubsubsubsection}{1em}{}
\titlespacing*{\subsubsubsection}
{0pt}{3.25ex plus 1ex minus .2ex}{1.5ex plus .2ex}

\makeatletter
\renewcommand\paragraph{\@startsection{paragraph}{5}{\z@}%
  {3.25ex \@plus1ex \@minus.2ex}%
  {-1em}%
  {\normalfont\normalsize\bfseries}}
\renewcommand\subparagraph{\@startsection{subparagraph}{6}{\parindent}%
  {3.25ex \@plus1ex \@minus .2ex}%
  {-1em}%
  {\normalfont\normalsize\bfseries}}
\def\toclevel@subsubsubsection{4}
\def\toclevel@paragraph{5}
\def\toclevel@paragraph{6}
\def\l@subsubsubsection{\@dottedtocline{4}{9em}{4em}}
\def\l@paragraph{\@dottedtocline{5}{10em}{5em}}
\def\l@subparagraph{\@dottedtocline{6}{14em}{6em}}
\makeatother

\usepackage{tikz}

\usetikzlibrary{arrows}
\usetikzlibrary{shapes}

\newcommand*\circled[1]{\tikz[baseline=(char.base)]{
            \node[shape=circle,draw,inner sep=2pt] (char) {#1};}}

\usepackage{pstricks,pst-node,pst-coil,pst-text,multido,pst-3d,pst-tree,pst-grad}

\usepackage{wrapfig}
\usepackage{colortbl}
\usepackage{etex}
\usepackage[ruled]{algorithm2e}
\usepackage{amsmath}
\usepackage{algorithmic}
\usepackage{amsfonts}
\usepackage{amsmath}
\usepackage{amssymb}
\usepackage{latexsym}
\usepackage{graphicx}
\usepackage{color}
\usepackage[normalem]{ulem}
\usepackage{balance}
\usepackage{textcomp}
\usepackage{nicefrac}
\usepackage{wasysym}
\usepackage{enumitem}
\usepackage{gensymb}
\usepackage{ctable}
\usepackage{url}
\usepackage{longtable}
\usepackage{multirow}
\usepackage{caption}

\usepackage{frcursive}
\usepackage{pifont,epsfig,rotating}
\usepackage{etoolbox}
\DeclareMathAlphabet{\mathpzc}{OT1}{pzc}{m}{it}
\usepackage{etaremune}

\definecolor{dgreyblue}{rgb}{0.26,0.3,0.46}             

\newcommand{\cI}{\mathcal{I}}

\newcommand{\cD}{\mathcal{D}}

\renewcommand{\text}[1]{\hbox{\rm \ #1\ \/}}

\newcommand{\be}[1]{\begin{equation}\label{#1}}
\newcommand{\ee}{\end{equation}}
\newcommand{\beqn}{\begin{eqnarray*}}
\newcommand{\eeqn}{\end{eqnarray*}}
\newcommand{\beq}{\begin{eqnarray}}
\newcommand{\eeq}{\end{eqnarray}}
\newcommand{\ben}{\begin{enumerate}}
\newcommand{\een}{\end{enumerate}}
\newcommand{\bi}{\begin{itemize}}
\newcommand{\ei}{\end{itemize}}

\newcommand{\cU}{\mathcal{U}}

\newcommand{\vecd}{\mathbf{d}}

\newcommand{\IE}{{\em i.e.}\xspace}
\newcommand{\tx}{^{\rm th}}

\newcommand{\refine}{ \prec_{\mathpzc{r}} }

\newtheorem{problem}{Problem}
\newtheorem{theorem}{Theorem}

\newtheorem{lemma}[theorem]{Lemma}

\newenvironment{proof-sketch}{{\noindent\bf Sketch of Proof.\ }}{\hfill{\Pisymbol{pzd}{113}}\vspace{0.1in}}

\newcommand{\NP}{\mathsf{NP}}

\newcommand{\nbr}{\mathsf{Nbr}}

\newcommand{\EA}{{\em et al.}\xspace}
\newcommand{\TB}{\vspace{-0.1ex}}\newcommand{\TiE}{\setlength{\itemsep}{-1ex}}

\newcommand{\comment}[1]{}

\newcommand{\EG}{{\it e.g.}\xspace}
\newcommand{\FI}[1]{Fig~\ref{#1}\xspace}

\newcommand{\apsp}{{\sc Apsp}\xspace}

\newcommand{\iif}{{\bf{if}}}
\newcommand{\tthen}{{\bf{then}}}
\newcommand{\eelse}{{\bf{else}}}
\newcommand{\ffor}{{\bf{for}}}
\newcommand{\wwhile}{{\bf{while}}}
\newcommand{\rreturn}{{\bf{return}}}

\newcommand{\ddo}{{\bf{do}}}

\newcommand{\dist}{\mathrm{dist}}

\newcommand{\kopt}{ {k}_{\mathrm{opt}} }
\newcommand{\popt}{ {p}_{\mathrm{opt}} }

\newcommand{\loptgt}{ \mathpzc{L}_{\mathrm{opt}}^{\geq k} }
\newcommand{\lopteq}{ \mathpzc{L}_{\mathrm{opt}}^{=k} }
\newcommand{\lopteqone}{ \mathpzc{L}_{\mathrm{opt}}^{=1} }

\newcommand{\hatlopteqone}{ \widehat{ \mathpzc{L}_{\mathrm{opt}}^{=1} } }
\newcommand{\hatloptgt}{ \widehat{ \mathpzc{L}_{\mathrm{opt}}^{\geq k} } }
\newcommand{\vopt}{ V_{\mathrm{opt}} }
\newcommand{\voptgt}{ V_{\mathrm{opt}}^{\geq k} }
\newcommand{\vopteq}{ V_{\mathrm{opt}}^{=k} }
\newcommand{\vopteqone}{ V_{\mathrm{opt}}^{=1} }

\newcommand{\hatvopteqone}{ \widehat{ V_{\mathrm{opt}}^{=1} } }
\newcommand{\hatvoptgt}{ \widehat{ V_{\mathrm{opt}}^{\geq k} } }

\newcommand{\nokmad}{{\sc Adim}}
\newcommand{\mad}{{\sc Adim}$_{\geq k}$}
\newcommand{\eqmad}{{\sc Adim}$_{= k}$}
\newcommand{\eqmadone}{{\sc Adim}$_{= 1}$}

\newcommand{\eqdef}{\stackrel{\mathrm{def}}{=}}
\definecolor{columbiablue}{rgb}{0.61, 0.87, 1.0}

\allowdisplaybreaks

\newcommand{\rom}[1]{\uppercase\expandafter{\romannumeral #1\relax}}

\usepackage[normalem]{ulem} 
\newcommand\hl{\bgroup\markoverwith
  {\textcolor{yellow}{\rule[-.5ex]{2pt}{2.5ex}}}\ULon}

\usepackage[bordercolor=white,backgroundcolor=gray!30,linecolor=black,colorinlistoftodos]{todonotes}

\renewcommand{\hl}[1]{#1}

\begin{document}

\begin{frontmatter}

\title{On analyzing and evaluating privacy measures for social networks under active attack} 

\author[hhh1]{Bhaskar DasGupta\corref{mycorrespondingauthor}\fnref{ggg1}}
\cortext[mycorrespondingauthor]{Corresponding author}
\ead{bdasgup@uic.edu}

\author[hhh1]{Nasim Mobasheri\fnref{ggg1}}
\ead{nmobas2@uic.edu}

\author[hhh2]{Ismael G. Yero\fnref{ggg2}}
\ead{ismael.gonzalez@uca.es}

\fntext[ggg1]{Research partially supported by NSF grant IIS-1160995.}

\fntext[ggg2]{This research was done while the author was visiting the University of Illinois at Chicago, USA, supported by 
``Ministerio de Educaci\'on, Cultura y Deporte'', Spain, under the 
``Jos\'e Castillejo'' program for young researchers (reference number: CAS15/00007)}

\address[hhh1]{Department of Computer Science, University of Illinois at Chicago, Chicago, IL 60607, USA}

\address[hhh2]{Departamento de Matem\'{a}ticas, Escuela Polit\'{e}cnica Superior, Universidad de C\'{a}diz, 11202 Algeciras, Spain}

\begin{abstract}
Widespread usage of complex interconnected social networks 
such as \emph{Facebook}, \emph{Twitter} and \emph{LinkedIn}
in modern internet era
has also unfortunately opened the door for privacy violation of users of such networks  
by malicious entities.
In this article we investigate, both theoretically and empirically, 
privacy violation measures of large networks under 
active attacks that was recently introduced in 
(\emph{Information Sciences}, 328, 403--417, 2016).
Our theoretical result indicates that the network manager responsible for prevention of 
privacy violation must be very careful in designing the network \emph{if its topology does not
contain a cycle}. 
Our empirical results shed light on privacy violation properties of eight real social networks 
as well as a large number of synthetic networks generated by both the 
classical Erd\"{o}s-R\'{e}nyi model and
the scale-free random networks generated by the 
Bar\'{a}basi-Albert preferential-attachment model.
\end{abstract}


\end{frontmatter}


\section{Introduction}

Due to a significant growth of applications of graph-theoretic methods to 
the field of social sciences in recent days,  
it is by now a standard practice to use the concepts and terminologies 
of network science to those social networks that focus on interconnections between people. 
However, social networks in general may represent much more than 
just networks of interconnections between people.
Rapid evolution of popular social networks such as 
\emph{Facebook}, \emph{Twitter} and \emph{LinkedIn} 
have rendered modern society heavily dependent on such virtual platforms for their day-to-day operation.
The powers and implications of social network analysis are indeed \emph{indisputable}; for example, 
such analysis may uncover previously unknown knowledge on community-based involvements, media usages and individual engagements. 
However, all these benefits are \emph{not} necessarily cost-free since 
a malicious individual could compromise privacy of users of these social networks for harmful purposes that may 
result in the disclosure of sensitive data (attributes) 
that may be linked to its users,  
such as node degrees, inter-node distances or network connectivity.
A natural way to avoid this consists of an ``anonymization process'' of the relevant social network in question. 
However, since such anonymization processes may \emph{not} always succeed, 
an important research goal is to be able to quantify and measure 
how much privacy a given social network can achieve. 
{Towards this goal, the recent work in~\cite{TY16} 
aimed at evaluating the \emph{resistance} of a social network against active privacy-violating attacks by 
introducing and studying theoretically 
a new and meaningful privacy measure for social networks}.
This privacy measure arises from the concept of the so-called $k$-metric antidimension of graphs that we explain next.

Given a connected simple graph $G = (V, E)$, and an ordered sequence of nodes 
$S = \left(v_{1}, \dots, v_{t} \right)$, the 
\emph{metric representation} of a node $u$ that is \emph{not} in $S$ with respect to $S$ is the vector (of $t$ components)
$\vecd_{u,-S} = (\dist_{u, v_1}, \dots, \dist_{u, v_t})$, 
where $\dist_{u, v}$ represents the length of a shortest path between nodes $u$ and $v$.
The set $S$ is then a $k$-\emph{antiresolving set} if $k$ is the largest positive integer such that for every node 
$v$ not in $S$ there also exist \emph{at least} other $k-1$ different nodes 
$v_{j_1},\dots, v_{j_{k-1}}$ not in $S$ such that $v,v_{j_1},\dots,v_{j_{k-1}}$ have the \emph{same} metric
representation with respect to $S$ (\IE, 
$\vecd_{v,-S} = \vecd_{v_{j_1},-S} = \dots = \vecd_{v_{j_{k-1}},-S}$). 
The $k$-\emph{metric antidimension} of $G$ is defined to be 
value of the minimum cardinality among all the $k$-antiresolving sets of $G$~\cite{TY16}.
If a set of attacker nodes $S$ represents a $k$-antiresolving set in a graph $G$, 
then an adversary controlling the nodes in $S$ cannot \emph{uniquely} re-identify
other nodes in the network (\emph{based on the metric representation}) 
with probability higher than $\nicefrac{1}{k}$. 
However, given that $S$ is unknown, any privacy measure for a social network should quantify over 
\emph{all} possible subsets $S$ of nodes. 
{\em
In this sense, a social network ${G}$ meets ${(k,\ell)}$-{anonymity} with respect
to active attacks to its privacy 
if ${k}$ is the smallest positive integer such that the ${k}$-metric antidimension of ${G}$ is no more than ${\ell}$.
In this definition of ${(k,\ell)}$-anonymity the parameter ${k}$ is used for a privacy threshold, 
while the parameter ${\ell}$ represents an {upper bound} on the expected number of attacker nodes in the network. 
}
Since attacker nodes are in general difficult to inject without being detected, the value
$\ell$ could be estimated based on some statistical analysis of other known networks. 
A simple example that explains the role of $k$ and $\ell$ to readers is as follows. 
Consider a complete network $K_n$ on $n$ nodes in which every node is connected with every other node.
It is readily seen that for any $0 < \ell < n$, this network meets $(n - \ell,\ell)$-anonymity.
In other words, this means that a social network $K_n$ guarantees that a user cannot be re-identified 
(based on the metric representation) 
with a probability higher than ${1}/({n - \ell})$ by an adversary controlling 
at most $\ell$ attacker nodes.
For other related concepts for metric dimension of graphs, the reader may consult references
such as~\cite{DM17,HSV12,KRR96}. 

Chatterjee \EA in~\cite{CDMVY16} (see also~\cite{ZG17}) 
formalized and analyzed the computational complexities of several optimization 
problems {motivated by the ${(k,\ell)}$-anonymity of a network as described in}~\cite{TY16}. 
In this article, we consider three of these optimization problems 
from~\cite{CDMVY16},
namely Problems~\ref{prob1}--\ref{prob3}
as defined in Section~\ref{sec-notation}.
A high-level itemized overview of the contribution of this article is as follows 
(see Section~\ref{sec-result} for precise technical statements and details of all contributions):
\begin{enumerate}[label=$\triangleright$]
\item
Our theoretical result concerning the anonymity issues for networks without cycles is provided in 
Theorem~\ref{thm1}
in Section~\ref{sec-result-theory}.
Some consequences of this theorem are also discussed 
\emph{immediately following a statement of the theorem}.
\item
In Section~\ref{sec-result-empirical},
we first describe briefly efficient implementations of 
the high-level algorithms of 
Chatterjee \EA~\cite{CDMVY16}
for Problems~\ref{prob1}--\ref{prob3}
(namely Algorithms I and II in Section~\ref{sec-alg-12}).
We then tabulate and discuss the results of applying these implemented algorithms for the following type of 
network data:
\begin{enumerate}[label=$\triangleright$]
\item 
eight real social networks listed in Table~\ref{L1}
in Section~\ref{sec-real-networks},
\item 
the classical undirected Erd\"{o}s-R\'{e}nyi random networks $G(n,p)$
for four suitable combinations of $n$ and $p$, and 
\item 
the \emph{scale-free random networks} $G(n,q)$ generated by the 
Bar\'{a}basi-Albert {\em preferential-attachment} model
for four suitable combinations of $n$ and $q$.
\end{enumerate}
{The ${6}$ tables that provide tabulations of the empirical results are Tables~\ref{L2}--\ref{L6} 
in Section~\ref{sec-result-empirical}
and the type of conclusions that one can draw from these tables are stated in 
the ${11}$ conclusions numbered 
{\LARGE\ding{172}}--{\circled{${11}$}} in the same section}.
Despite our best efforts, we do not know of any other alternate approaches (\EG, sybil attack framework) 
that will provide a significantly simpler theoretical framework
to reach all the ${11}$ conclusions as mentioned above.
\end{enumerate}
As an illustration of a potential application, consider 
the \emph{hub fingerprint query} model of Hey \EA~\cite{Hay08}.
Noting that the largest hub fingerprint for a target node $u$ is the metric representation of $u$ with respect to the hub nodes,
results on $(k,\ell)$-anonymity are directly applicable to this setting of 
Hey \EA~\cite{Hay08} that models an adversary trying to identify the hub nodes in a network.
For example, assuming that the quantity $\kopt$ in Problem~\ref{prob1} (see Section~\ref{sec-notation} for a definition)
is $10$, the network is vulnerable with respect to hub identification in the model of Hey \EA in the sense that 
it is \emph{not} possible to guarantee that an adversary 
will not be able to uniquely re-identify any node 
in the network with probability at most $0.1$.

\subsection{Some remarks regarding the model and our contribution (to avoid possible confusion)}

To avoid any possible misgivings or confusions regarding the technical content of the paper as well as 
to help the reader towards understanding the remaining content of this article, we believe the following 
comments and explanations may be relevant.
\begin{enumerate}[label=$\blacktriangleright$,leftmargin=*]
\item
{The computational complexity investigations in this paper has nothing to do with the model in 
the paper by Backstrom \EA~\cite{Backstrom}.}
{We \emph{whole-heartedly} and \emph{without any reservations} agree that the paper by 
Backstrom \EA~\cite{Backstrom} is seminal, but 
the research investigations in this paper has nothing to do with the model or any measure introduced
in the paper by Backstrom \EA~\cite{Backstrom}}.
The notion of active attack is very different in that paper, and therefore
{the computational problems that arise in that paper are very different from those in the current paper and 
in fact incomparable}.
{Finally, the goal of this paper is not to compare various network privacy models but to 
investigate, theoretically and empirically, the model in~\cite{TY16}}.
\item
{This paper does \emph{not} introduce any new privacy model or measure, but simply investigates, 
both theoretically and empirically, computational 
problems for a  model that is published in} ``\emph{Information Sciences}, 328, 403--417, 2016''
(reference~\cite{TY16}).
{There have been several other subsequent papers
investigating this privacy measure, \EG, see~\cite{CDMVY16,TY16-TCJ,ZG17,H}.
Thus, researchers in network privacy are certainly interested in this model
or related computational complexity questions}.
Of course, this does not contradict the fact that 
the paper by Backstrom \EA~\cite{Backstrom} is seminal.
\item
Even though the network privacy model was introduced in~\cite{TY16} and therefore the best option for clarification 
of any confusion regarding the model would be to look at that paper, 
we provide the following clarification just in case.
{In this model, nobody is trying to prevent adversaries.
Informally, the privacy measure only gives a ``measure'' on how much secure a 
graph is against active attacks, \IE, a probability with which we can assert that, if there are 
controlled nodes in a graph, then we can in some sense know which is the probability to be 
reidentified in such graph (for details please see the texts preceding and following the statements of 
Problems~\ref{prob1}--\ref{prob3} in Section~\ref{sec-notation}).
No new nodes are added at all. This is not a problem that involves dynamic graphs.
The model in~\cite{TY16} is not the same as the one by Backstrom \EA~\cite{Backstrom}}.
\end{enumerate}

\subsection{Comparison with other existing works}

\paragraph{Model comparison}
Unfortunately, different models of network privacy have quite different objectives and consequently 
quite different measures that cannot in general be compared to one another.
In particular, we know of no other different but comparable model or measure of network privacy
that can be compared to those in our paper.
For example, the network privacy model introduced by  
Backstrom \EA~\cite{Backstrom} is interesting, but 
the notion of active attack is very different in that paper, and therefore
the computational problems that arise in that paper are very different from those in the current paper and 
in fact \emph{incomparable}.

\paragraph{Algorithmic comparison}
Note that algorithms for different models \emph{cannot} be compared in terms of their worst-case 
(or average-case) computational complexities.
For example, 
consider the \emph{scale-free network model} and the computational complexity 
paper for this model in~\cite{GHK15}.
Now, consider the \emph{Erd\"{o}s-R\'{e}nyi random regular network} model, and consider the paper
in~\cite{Z01}.
Even though~\cite{Z01} provides better algorithmic results in terms of time-complexity and approximability, 
that does not nullify the research results 
in~\cite{GHK15}.


\paragraph{Privacy preservation in learning theoretic framework}
The recent surge in popularity of machine learning applications to different domains, specifically
in the context of \emph{deep learning} methods, has motivated many Internet companies to 
provide numerous online cloud-based services and frameworks for developing and deploying machine learning applications
(Machine Learning as a Service or MLaaS)
such as the Google Cloud ML Engine.
Typically, an user (customer) of such a system first estimates the parameters of a suitable model by training the model
with data and afterwards, once the correct model is determined, uploads the model to the cloud provider
such that remote users can use the model. This type of service frameworks lead to two possible privacy concerns, the first
concerning privacy violations of the training data, and the second concerning privacy violations of data uploaded 
by remote users. For some recent papers dealing with possible remedies of these privacy violations, such as 
introducing suitable random noises to perturb the data, see papers such as~\cite{Salem,Zhang}.
However, these privacy concerns are quite different from the current topic of our paper, such as they are not 
specific to networks and they involve learning paradigms which are not of interest to this paper. 
Whether privacy questions in the MLaaS framework can be combined with those in this paper is an 
interesting research question but unfortunately beyond the scope of this paper.

\section{Basic notations, relevant background and problem formulations}
\label{sec-notation}

\begin{figure}[!h]
\includegraphics[scale=0.9]{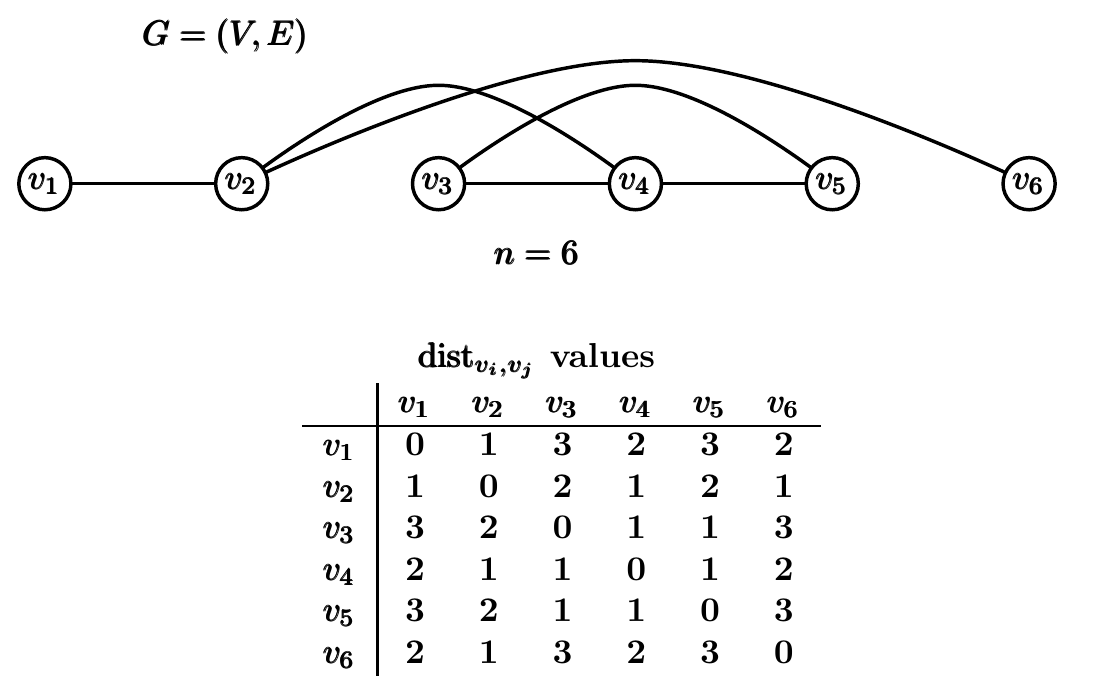}
\caption{\label{fig0}An example for illustration of some basic definitions and notations in Section~\ref{sec-notation}.}
\end{figure}

Let $G=(V,E)$ be the undirected input network over $n$ nodes $v_1,\dots,v_n$. 
The authors in~\cite{CDMVY16} formalized and analyzed the computational complexities of several optimization 
problems motivated by the $(k,\ell)$-anonymity of a network as described in~\cite{TY16}.
The notations and terminologies from~\cite{CDMVY16} relevant for this paper are as follows ({\em see \FI{fig0} for 
an illustration})\footnote{The notations and the theoretical frameworks are actually \emph{not} that complicated
once one goes over them carefully. Although one may wonder
if significantly simpler notations could have been adopted without neglecting the complexities of the frameworks, 
it does not seem to be possible in spite of our best efforts for over an year.}: 
\begin{enumerate}[label=$\blacktriangleright$,leftmargin=*]
\item 
$\vecd_{v_i}=\left(\dist_{v_i,v_1},\dist_{v_i,v_2},\dots,\dist_{v_i,v_n}\right)$
denotes the metric representation of a node $v_i$.
For example, 
in \FI{fig0},  
$\vecd_{v_1}=\left(0,1,3,2,3,2\right)$.
\item 
$\nbr\left(v_\ell\right)=
\left\{ \, v_j \,|\, \left\{v_\ell,v_j\right\} \in E \, \right\}$
is the (open) \emph{neighborhood} of node $v_\ell$ in $G=(V,E)$.
For example, 
in \FI{fig0},
$\nbr\left(v_2\right)=
\left\{ \, v_1, v_4, v_6 \, \right\}$.
\item 
For a subset of nodes $V'=\left\{ v_{j_1},v_{j_2},\dots,v_{j_t}\right\}\subset V$ with 
$j_1<j_2<\dots<j_t$ and any other node $v_i\in V\setminus V'$,
$\vecd_{v_i,-V'} = \big( \dist_{v_i,v_{j_1}}, \dist_{v_i,v_{j_2}}, \dots, \dist_{v_i,v_{j_t}} \big)$
denotes the metric representation 
of $v_i$ with respect to $V'$.
The notation is further generalized by defining 
$\cD_{V'',-V'}=\left\{ \, \vecd_{v_i,-V'} \,|\, v_i\in V'' \, \right\}$
for any $V'' \subseteq V\setminus V'$.
For example, 
in \FI{fig0},
$\vecd_{v_3,-\left\{v_1,v_5,v_6 \right\} } = \big( \underset{v_1}{3}, \underset{v_5}{1}, \underset{v_6}{3}\big)$
and 
$\cD_{\left\{ v_2, v_3 \right\},-\left\{v_1,v_5,v_6 \right\} } = 
\big\{ ( \overbrace{ \underset{v_1}{1},\underset{v_5}{2},\underset{v_6}{1} }^{\text{\footnotesize from $v_2$}} ), 
( \overbrace{ \underset{v_1}{3}, \underset{v_5}{1}, \underset{v_6}{3} }^{\text{\footnotesize from $v_3$}} ) \big\}$.
\item 
A partition $\Pi'=\left\{ V_1',V_2',\dots,V_\ell' \right\}$ 
of $S'\subseteq V$ 
is called a \emph{refinement}
of 
a partition 
$\Pi=\left\{ V_1,V_2,\dots,V_k \right\}$ 
of $S\supseteq S'$,  
denoted by $\Pi'\refine\Pi$, provided
$\Pi'$ can be obtained from $\Pi$ in the following manner:
\begin{enumerate}[label=$\triangleright$,leftmargin=*]
\item 
For every node $v_i\in \left( \cup_{t=1}^k V_t \right)
\setminus \left( \cup_{t=1}^\ell V'_t \right)
$, remove $v_i$ from the set in $\Pi$ that contains it.
\item 
\emph{Optionally}, for every set $V_\ell$ in $\Pi$, replace $V_\ell$ by a partition of $V_\ell$.
\item 
Remove empty sets, if any.
\end{enumerate}
For example, 
for \FI{fig0},  
$
\left\{ 
\{v_2\},\{v_3\},\{v_4,v_5\}
\right\}
\refine
\left\{ 
\{v_1,v_2,v_3\},\{v_4,v_5\}
\right\}
$.
\item 
The following notations pertain to the 
equality relation (an equivalence relation) 
over the set of (same length) vectors $\cD_{V\setminus V',-V'}$ for some $\emptyset\subset V'\subset V$: 
\begin{enumerate}[label=$\triangleright$,leftmargin=*]
\item 
The set of equivalence classes, which forms a partition of $\cD_{V\setminus V',-V'}$, is denoted by
$\Pi_{{V\setminus V',-V'}}^{=}$.
For example, 
in \FI{fig0},  
$
\cD_{\left\{ v_2, v_3, v_4, v_5 \right\},-\left\{v_1, v_6 \right\} } = 
\big\{ 
( \overbrace{ \underset{v_1}{1},\underset{v_6}{1} }^{\text{\footnotesize from $v_2$}} ), 
( \overbrace{ \underset{v_1}{3},\underset{v_6}{3} }^{\text{\footnotesize from $v_3$}} ), 
( \overbrace{ \underset{v_1}{2},\underset{v_6}{2} }^{\text{\footnotesize from $v_4$}} ), 
( \overbrace{ \underset{v_1}{3},\underset{v_6}{3} }^{\text{\footnotesize from $v_5$}} )
\big\}$
and 
\\
$
\Pi_{\left\{ v_2, v_3, v_4, v_5 \right\},-\left\{v_1, v_6 \right\} }^{=} = 
\Big\{ 
\,
\big\{ 
( \overbrace{ \underset{v_1}{1},\underset{v_6}{1} }^{\text{\footnotesize from $v_2$}} )
\big\} , 
\big\{ 
( \overbrace{ \underset{v_1}{2},\underset{v_6}{2} }^{\text{\footnotesize from $v_4$}} )
\big\} , 
\big\{ 
( \overbrace{ \underset{v_1}{3},\underset{v_6}{3} }^{\text{\footnotesize from $v_3$}} ), 
( \overbrace{ \underset{v_1}{3},\underset{v_6}{3} }^{\text{\footnotesize from $v_5$}} )
\big\}
\,
\Big\}$.
\item 
Abusing terminologies slightly, two nodes $v_i,v_j\in V\setminus V'$ will be said to
belong to the \emph{same} equivalence class if $\vecd_{v_i,-V'}$ and
$\vecd_{v_j,-V'}$ belong to the same equivalence class in
$\Pi_{{V\setminus V',-V'}}^{=}$, and thus
$\Pi_{{V\setminus V',-V'}}^{=}$
also defines a partition into equivalence classes of $V\setminus V'$.
For example, 
in \FI{fig0},  
$v_3$ and $v_5$
belong to the same equivalence class in 
$
\Pi_{\left\{ v_2, v_3, v_4, v_5 \right\},-\left\{v_1, v_6 \right\} }^{=}
$ 
and 
$
\Pi_{\left\{ v_2, v_3, v_4, v_5 \right\},-\left\{v_1, v_6 \right\} }^{=}
$ 
also defines the partition 
$
\big\{
\{v_2\}, 
\{v_4\}, 
\{v_3,v_5\}
\big\}
$.
\item 
The \emph{measure} of the equivalence relation is defined as
$
\mu \left(\cD_{V\setminus V',-V'}\right)
\eqdef
\min_{\mathcal{Y}\in\Pi_{{V\setminus V',-V'}}^{=}}
\big\{
\,\left|
\,\mathcal{Y}\,
\,\right|
\big\}
$.
Thus, if a set $S$ is a $k$-antiresolving set, then
$\cD_{V\setminus S,-S}$ defines a partition into equivalence classes whose measure
is $k$.
For example, 
in \FI{fig0},  
$\mu \left( \Pi_{\left\{ v_2, v_3, v_4, v_5 \right\},-\left\{v_1, v_6 \right\} }^{=} \right) =1$.
\end{enumerate}
\end{enumerate}
%
%
%
By using the terminologies mentioned above, the following three optimization problems were formalized
and studied in~\cite{CDMVY16}.
\emph{
We need to stress that one really needs to study the three different problems and 
consequently the three objectives (namely, ${\kopt}$, ${\loptgt}$ and ${\lopteq}$)
separately because they are motivated by different considerations as explained before and after
the problem definitions and as stated in (${\star}$), (${\bowtie}$) and (${\spadesuit}$).
Informally and briefly, 
Problem~\ref{prob1} and ${\kopt}$ 
are used to provide an absolute privacy violation bound assuming the attacker can control as
many nodes as it needs, 
restricting the number of attacker
nodes employed by the adversary 
leads to Problem~\ref{prob2}, and 
Problem~\ref{prob3} is motivated by a
type of trade-off question 
between ${(k,\ell)}$-anonymity vs.\ ${(k',\ell')}$-anonymity.
Thus, it is simply not possible to combine them into fewer than three problems.
}

\begin{problem}[metric anti-dimension or \nokmad)]\label{prob1}
Find a subset of nodes $V'$ such that
$
\kopt =
\mu \left(\cD_{V\setminus V',-V'} \right)
=
\max\limits_{\emptyset \subset S\subset V}
\big\{
\,
\mu \left(  \cD_{V\setminus S,-S} \right)
\big\}
$.
\end{problem}

A solution of Problem~\ref{prob1} asserts the following:

\begin{quote}
\begin{description}
\item[($\star$)] 
Assuming that there is \emph{no} restriction on the number of nodes that can be controlled by an adversary,
the following statements hold:
\begin{description}
\item[(\emph{a})]
The network administrator \emph{cannot} guarantee that an adversary 
will not be able to uniquely re-identify any node 
in the network (based on the metric representation) 
with probability $\nicefrac{1}{\kopt}$ or less.
\item[(\emph{b})]
It \emph{is} possible for an adversary to uniquely re-identify
$\kopt$ nodes 
in the network (based on the metric representation) 
with probability $\nicefrac{1}{\kopt}$.
\end{description}
\end{description}
\end{quote}

Thus, informally, 
Problem~\ref{prob1} 
and $\kopt$ 
give an absolute privacy violation bound assuming the attacker can control as
many nodes as it needs.
In practice, however, the number of attacker
nodes employed by the adversary \emph{may} be restricted.
This leads us to Problem~\ref{prob2}.

\begin{problem}[$k_\geq$-metric anti-dimension or \mad]\label{prob2}
Given a positive integer $k$,
find a subset 
$\voptgt$ 
of nodes 
of \emph{minimum} cardinality
$\loptgt=\big| \voptgt \big|$, 
if one such subset at all exists, such that
$\mu \big(  \cD_{V\setminus \voptgt,-\voptgt} \big)\geq k$.
\end{problem}

Similar to ($\star$), 
a solution of Problem~\ref{prob2} (if it exists) asserts the following: 

\begin{quote}
\begin{description}
\item[($\bowtie$)] 
Assuming that an adversary may control up to $\alpha$ nodes,
the following statements hold:
\begin{description}
\item[(\emph{a})]
If $\alpha<\loptgt$ 
then 
the network administrator \emph{can} guarantee that an adversary 
will not be able to uniquely re-identify any node 
in the network (based on the metric representation) 
with probability $\nicefrac{1}{k}$ or less.
\item[(\emph{b})]
If $\alpha\geq\loptgt$ 
then 
the network administrator \emph{cannot} guarantee that an adversary 
will not be able to uniquely re-identify any node 
in the network (based on the metric representation) 
with probability $\nicefrac{1}{k}$ or less.
\item[(\emph{c})]
If $\alpha\geq\loptgt$ 
then 
it \emph{is} possible for an adversary to uniquely re-identify
a subset of $\beta$ nodes 
in the network (based on the metric representation) 
with probability $\nicefrac{1}{\beta}$
for some $\beta\geq k$
(note that $\beta$ may be much larger compared to $k$).
\end{description}
\end{description}
\end{quote}

The remaining third problem is motivated by the following 
trade-off question 
between $(k,\ell)$-anonymity vs. $(k',\ell')$-anonymity:
if $k'>k$ but $\ell'<\ell$ then 
$(k',\ell')$-anonymity
has \emph{smaller} privacy 
violation probability 
$\nicefrac{1}{k'}<\nicefrac{1}{k}$
compared to 
$(k,\ell)$-anonymity
but can only tolerate attack on 
\emph{fewer} 
$\ell'<\ell$
number 
of nodes.

\begin{problem}[$k_{=}$-metric antidimension or \eqmad]\label{prob3}
Given a positive integer $k$,
find a subset 
$\vopteq$ 
of nodes 
of \emph{minimum} cardinality 
$\lopteq=\big| \vopteq \big|$, 
if one such subset at all exists, such that
$\mu \big(  \cD_{V\setminus \vopteq,-\vopteq} \big)=k$.
\end{problem}

One can describe assertions to
a solution of Problem~\ref{prob2} (if it exists) in a manner similar to that in 
($\star$)
and
($\bowtie$).
Chatterjee \EA in~\cite{CDMVY16} studied the computational 
complexity aspects of Problems~\ref{prob1}--\ref{prob3}.
They provided efficient (polynomial-time) algorithms to solve 
Problems~\ref{prob1}~and~\ref{prob2}
and showed that Problem~\ref{prob3} is \emph{provably} computationally hard for exact solution but admits
an efficient approximation for 
the particular case of $k=1$ (see Algorithm~II). 
Since we use this approximation algorithm for $k=1$, we explicitly state below the 
implication of a solution of \eqmadone\ (note that a solution of \eqmadone\ always exists and 
$\lopteqone$ is trivially at most $n-1$):

\begin{quote}
\begin{description}
\item[($\spadesuit$)] 
It suffices for an adversary to control 
a \emph{suitable} subset of $\lopteqone$ nodes in the network to
\emph{uniquely} re-identify
at least one node in the network (based on the metric representation) 
with \emph{absolute certainty} (\IE, with a probability of one).
\end{description}
\end{quote}

\section{Our theoretical and empirical results}
\label{sec-result}

\subsection{Theoretical result}
\label{sec-result-theory}

Suppose that a given graph $G$ is a ``$k'$-metric antidimensional'' graph, \IE, 
$k'$ is the largest positive integer such that $G$ has \emph{at least} one
$k'$-antiresolving set.
Then obviously $G$ does 
\emph{not} contain any $k''$-antiresolving set for every $k''> k'$. 
In contrast, it is not \emph{a priori} clear if $G$ contains $k$-metric antiresolving sets for any $k<k'$. 
For instance, a complete graph $K_n$ on $n$ nodes is $(n-1)$-metric antidimensional and  moreover, 
for every $1\le k\le n-1$, there exists a set of nodes in $K_n$ which is a $k$-antiresolving set. 
\emph{Au contraire}, 
if we consider the wheel graph
$W_{1,n}$ (see \FI{fig1} for an illustration for $n=16$), 
it is easy to see that the central node $v_n$ is the \emph{unique} $n$-antiresolving set,
$1$-antiresolving and $2$-antiresolving sets exist, $3$-antiresolving sets also exist (if $n$ is larger than $5$), but \emph{no} 
$k$-antiresolving set exists for $4\le k\le n-1$. 
This motivates the following research question:

\begin{quote}
{\em For a given class of $k'$-metric antidimensional networks, can we decide if they also have  
$k$-antiresolving sets for all $1\le k\le k'-1$}?
\end{quote}

\begin{figure}[!h]
\centerline{\includegraphics{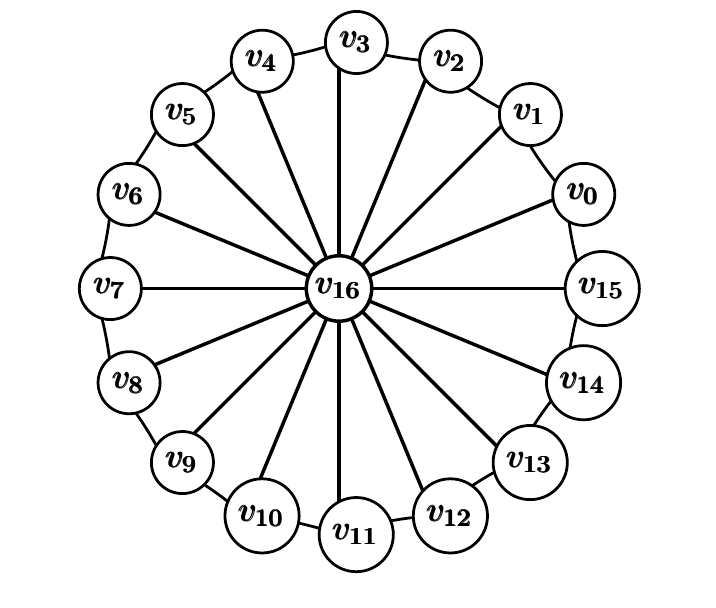}}
\caption{The wheel graph $W_{1,n}$ for $n=16$.}
\label{fig1}
\end{figure}

The following theorem answers the question affirmatively for all networks without a cycle.

\begin{theorem}\label{thm1}
If $T$ is a $k'$-metric antidimensional tree, then for every $1\le k\le k'$ there exists a $k$-antiresolving set for $T$.
\end{theorem}

\subsubsection*{Some consequences of Theorem~\ref{thm1}}

Some consequences of the above result in relation to the $(k,\ell)$-anonymity measure are as follows.
%
Note 
that what is stated below is \emph{not} the same as the observations in~\cite{H}.

Clearly, since trees have nodes of degree one (called leaves), it is always possible to identify at 
least one node of the tree~\cite{H}. 
However, if the network manager introduces some ``fake'' nodes as leaves, then this advantage for the adversary is avoided. 
In this sense, the result above asserts that an adversary will never be sure that the set of nodes which it could control 
will always identify at least one node of the given tree. Another related interesting observation is that for this to happen, 
the tree must be $k$-metric antidimensional for some $k\ge 2$, otherwise the tree is \emph{completely insecure}. 
A characterization of that trees which are $1$-metric antidimensional (graphs that contain only $1$-antiresolving sets) was given 
in~\cite{TY16-TCJ}.

\emph{Note that in the above we claim nothing about what happens if the network does contain a cycle, or how a network manager 
can break cycles in a network}.
Note that the topology need not be ``fully'' controlled by a network manager, but can be influenced by adding extra nodes.

\subsubsection*{Proof of Theorem~\ref{thm1}}

We will use the following result from~\cite{TY16-TCJ} in our proof.

\begin{lemma}{$\!\!\!$\em\cite{TY16-TCJ}}\label{lem_connected}
Any $k$-antiresolving set $S$ in a tree $T$ with $k\ge 2$ induces a connected subgraph of $T$.
\end{lemma}

Since Problem~\ref{prob1} was shown to be solvable in polynomial time in~\cite{CDMVY16}, we may assume that 
we know the value $k'$ for which the tree $T$ is $k'$-metric antidimensional.
If $k=1$ or $k=k'$ then a $k$-antiresolving set for $T$ clearly exists. We may also assume $k>1$, since otherwise our 
result follows trivially. Suppose that $k=k'-1$ and let $S$ be a $k'$-antiresolving set of minimum cardinality for $T$. By 
Lemma~\ref{lem_connected}, $S$ induces a connected subgraph of $T$. Moreover, according to the definition of a 
$k$-antiresolving set, there exists an equivalence class $Q\in\Pi_{{V\setminus S,-S}}^{=}$ 
such that $|Q|=k'$. Select $v\in S$ such that 
$\nbr(v)\setminus S\ne \emptyset$ and let $v_1,v_2,\dots,v_r\in \nbr(v)\setminus S$ for some $r\geq 1$. 
Clearly, the set $A_1=\left\{v_1,v_2,\dots,v_r\right\}$ forms an equivalence class of
$\Pi_{{V\setminus S,-S}}^{=}$. 
Moreover, the set $A_2=\bigcup_{i=1}^r \nbr(v_i)\setminus\{v\}$, if not empty,
also forms an equivalence class
of 
$\Pi_{{V\setminus S,-S}}^{=}$. 
\FI{trees-proof} shows two examples which are useful to clarify all the notations of this proof (recall that the 
\emph{eccentricity} of a node $v$ is the maximum over the set of
distances between $v$ to all other nodes in the graph).

\begin{figure}[!h]
\includegraphics[scale=0.7]{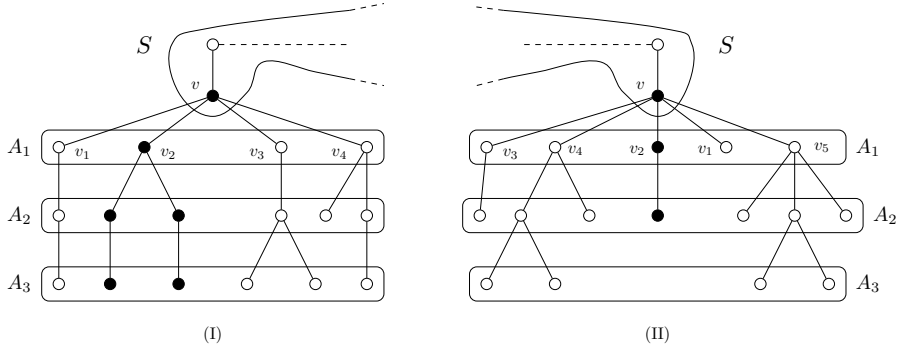}
\caption{Two auxiliary trees. Notice that eccentricity of $v$ in the subtrees is three in both cases. 
The set $S$ is a $4$-antiresolving set. The nodes of the subtree $T_2$ are shown in bold in both trees.}
\label{trees-proof}
\end{figure}

Assume that $T$ is rooted at node $v$ and, for every $v_i\in A_1$, let $T_i$ be the subtree of $T$ with node set 
$V(T_i)$ formed by $v$, $v_i$, 
and the set of descendants of $v_i$. Let $e_i$ be the eccentricity of $v$ in $T_i$ for $1\le i\le r$. 
Moreover, let $A_j$ be the subset of nodes $x$ in $\bigcup_{i=1}^{\,r} V(T_i)$ such that 
$\dist_{v,x}=j$ for every $1\le j\le \max\{e_i\,:\; 1\le i\le r\}$. 
Observe that each $A_j$, with $1\le j\le \max\{e_i\,:\; 1\le i\le r\}$, is an equivalence class of 
$\Pi_{{V\setminus S,-S}}^{=}$
and thus, $|A_j|\ge k'$ since otherwise $S$ is \emph{not} a $k'$-antiresolving set. 
Moreover, without loss of generality, we can assume there exists a set 
$A_q$ such that $|A_q|=k'$ (\EG, in \FI{trees-proof} the sets $A_1$ and $A_4$). If there is no such set, 
then we choose another node $v'$ of $T$ for which this situation happen. If there is no such node $v'$ at all,
then the cardinality of every equivalence class of 
$\Pi_{{V\setminus S,-S}}^{=}$
is \emph{strictly} larger than $k'$, 
which contradicts the definition of a $k'$-antiresolving set. We now consider the following situations.

\medskip

\noindent 
\textbf{Case 1: $e_1=e_2=\dots=e_r$} (\EG, in \FI{trees-proof}(I) all the eccentricities are equal to $3$). 
Notice that in this case 
$A_j\cap V(T_i)\ne \emptyset$
for every $1\le j\le \max\{e_i\,:\; 1\le i\le r\}$ and every $1\le i\le r$.
Moreover, there exist $\alpha,\beta$ such that $|A_{\alpha}\cap V(T_{\beta})|=1$ (\EG, in \FI{trees-proof}(I) 
$\alpha=1$ and $\beta$ can take any value between $1$ and $4$). 
Thus, for the set $S'=S\cup V(T_{\beta})$ it follows that $A_{\alpha}\setminus V(T_{\beta})$ 
is an equivalence class of the
equivalence relation 
$\Pi_{{V\setminus S',-S'}}^{=}$
and $|A_{\alpha}-V(T_{\beta})|=k'-1$. Moreover, for every other equivalence class $X$ of 
$\Pi_{{V\setminus S',-S'}}^{=}$
it follows $|X|\ge k'-1=k$. Thus, $X$ is a $(k'-1)$-antiresolving set. Clearly, $X$ could not be of minimum cardinality.

\medskip

\noindent 
\textbf{Case 2: There are at least two subtrees $T_i$ and $T_j$ such that $e_i\ne e_j$}. 
Without loss of generality, assume that 
$e_1\le e_2\le \dots \le e_r$. As in Case~1, there exist $\gamma$ such that $|A_{\gamma}|=k'$ (\EG, in \FI{trees-proof}(II) $\alpha=3$). 
Let $S_1=S\cup V(T_{1})$ (note that $T_1$ is the subtree in which $v$ has the minimum eccentricity). 
If $|A^{(1)}_j|\ge k'$
for every $A^{(1)}_j=A_j\setminus V(T_1)$ with $1\le j\le e_1$,  
then $\gamma > e_1$ and thus $S_1$ is \emph{also} a $k'$-antiresolving set. 
Hence, we consider $S_2=S_1\cup V(T_{2})$ (note that $T_2$ is the subtree in which $v$ has the second minimum eccentricity). 
If $|A^{(2)}_j|\ge k'$ for every $A^{(2)}_j=A^{(1)}_j\setminus V(T_2)$ with $1\le j\le e_2$, 
then $\gamma > e_2$. Repeating this procedure, we shall find a set 
$S_q=S_{q-1}\cup V(T_{q})$ such that $\gamma\le e_q$ and moreover, 
$|A_{\alpha'}\cap V(T_{\beta'})|=1$ for some $1\le \alpha'\le e_r$ and $q\le \beta'\le r$. 
Thus, the set $A^{(q+1)}_j=A^{(q)}_j\setminus V(T_{q+1})$ satisfies $|A^{(q)}_j|=k'-1$ and consequently
$S_{q+1}=S_q\cup V(T_{q+1})$ is a $(k'-1)$-antiresolving set (\EG, in \FI{trees-proof}(II) the process must be done two times, 
first we remove the nodes in the set $V(T_1)\setminus\{v\}$ and next we remove the nodes in the set $V(T_2)\setminus\{v\}$,
thereby getting the required $(k'-1)$-antiresolving set).

\medskip

Thus, in both cases we obtain a $(k'-1)$-antiresolving set. By using the same procedure and a 
$(k'-1)$-antiresolving set of minimum cardinality, we can find a $(k'-2)$-antiresolving set and in general 
a $k$-antiresolving set for every $2\le k\le k'-1$, which completes the proof.

\subsection{Empirical results}
\label{sec-result-empirical}

We remind the readers about the assertions in ($\star$), ($\bowtie$) and ($\spadesuit$)
while we report our empirical results and related conclusions.

\subsubsection{Algorithms for Problems~1--3 (Algorithms~I~and~II)}
\label{sec-alg-12}

We obtain an exact solution for Problem~\ref{prob2}
by implementing the following algorithm (Algorithm~I) 
devised in~\cite{CDMVY16} 
by Chatterjee \EA.
In this algorithm, an absence of a valid solution is indicated by 
$\loptgt\leftarrow\infty$ and $\voptgt\leftarrow\emptyset$.

\smallskip
\begin{longtable}{r l l }
%
\multicolumn{3}{l}{
(* Algorithm I *) 
}
\\
{\bf 1.} &
\multicolumn{2}{l}{
Compute $\vecd_{v_i}$ for all $i=1,\dots,n$ 
using any algorithm that solves 
}
\\
&
\multicolumn{2}{l}{
\hspace*{0.2in}
\emph{all-pairs-shortest-path} problem~\cite{CLR90}.
}
\\
{\bf 2.} &
\multicolumn{2}{l}{
$\hatloptgt\leftarrow\infty$ ;
$\hatvoptgt\leftarrow\emptyset$
}
\\
{\bf 3.} &
\multicolumn{2}{l}{
\ffor\ each $v_i\in V$ \ddo
}
\\
{\bf 3.1} & &
  $V'=\left\{ v_i \right\}$ ; $\mathsf{done}\leftarrow\mathsf{FALSE}$
\\
{\bf 3.2} & &
\wwhile\ $\big($ $(V\setminus V'\neq\emptyset)$ \textsf{AND} $(\mathsf{NOT}\,\,\mathsf{done})$ $\big)$ \ddo
\\
{\bf 3.2.1}
& &
\hspace*{0.2in}
compute
$\mu \left(  \cD_{V\setminus V',-V'} \right)$
\\
{\bf 3.2.2}
& &
\hspace*{0.2in}
\iif\
$\Big( \, \big( \, \mu \left(  \cD_{V\setminus V',-V'} \right)\geq k \, \big)$ and
$\big( \, |V'|<\hatloptgt \, \big) \, \Big)$
\\
{\bf 3.2.3}
& &
\hspace*{0.2in}
$\,\,\,\,\,$
\tthen\
$\,\,\,\,\,$
$\hatloptgt\leftarrow|V'|$ ;
$\hatvoptgt\leftarrow V'$ ;
$\mathsf{done}\leftarrow\mathsf{TRUE}$
\\
{\bf 3.2.4}
& &
\hspace*{0.4in}
\eelse\
$\,\,\,\,\,\,\,\,$
let $V_1,V_2,\dots,V_\ell$ be the \emph{only} $\ell>0$ equivalence classes 
\\
& &
\hspace*{1.1in}
in $\Pi_{{V\setminus V',-V'}}^{=}$
such that
\\
& &
\hspace*{1.1in}
$\left|V_1\right|=\dots=\left|V_\ell\right|=\mu \left( \cD_{V\setminus V',-V'} \right)$
\\
{\bf 3.2.5}
& &
\hspace*{0.95in}
$V'\,\leftarrow\,V'\cup \left( \cup_{t=1}^\ell V_t \right)$
\\
{\bf 4.} &
\multicolumn{2}{l}{
\rreturn\ $\hatloptgt$ and $\hatvoptgt$ as our solution
}
\end{longtable}
\smallskip

We obtain exact solutions for Problem~\ref{prob1} and find $\kopt$ by using 
Algorithm~I 
and doing a binary search for the parameter $k$ over the range $\{1,2,\dots,n\}$ 
to find the largest $k$ such that 
$\voptgt\neq\emptyset$. This requires using
Algorithm~I $O(\log n)$ times.


%

Although \eqmad\ is $\NP$-hard for almost all $k$, for $k=1$ 
we implement the following 
logarithmic-approximation algorithm devised in~\cite{CDMVY16} 
by Chatterjee \EA
for \eqmadone\ computing $\lopteqone$ and $\vopteqone$.

\smallskip
\begin{longtable}{l l l }
\multicolumn{3}{l}{(* Algorithm II *)}
%
\\
%
{\bf 1.} &
\multicolumn{2}{l}{
Compute $\vecd_{v_i}$ for all $i=1,\dots,n$ 
using any algorithm that solves 
}
\\
&
\multicolumn{2}{l}{
\hspace*{0.2in}
\emph{all-pairs-shortest-path} problem~\cite{CLR90}.
}
\\
{\bf 2.} &
\multicolumn{2}{l}{
$\hatlopteqone\leftarrow\infty$ ;
$\hatvopteqone\leftarrow\emptyset$
}
\\
{\bf 3.} &
\multicolumn{2}{l}{
\ffor\ each node $v_i\in V$ \ddo
}
\\
& {\bf 3.1} &
      create the following instance of the set-cover problem~\cite{J74} 
\\
&           & 
			containing $n-1$ elements and $n-1$ sets:
\\
& &
	 \hspace*{0.15in}
   $\cU=\left\{ \,a_{v_j} \,|\, v_j\in V\setminus \left\{v_i\right\} \, \right\}$,
\\
& &
	 \hspace*{0.15in}
	 $S_{v_j}= \left\{ a_{v_j} \right\} \cup \left\{ a_{v_\ell} | \dist_{v_i,v_j}\neq\dist_{v_\ell,v_j}\right\}$
	        for $j\in \{1,\dots,n\}\setminus \{i\}$
\\
& {\bf 3.2} &
\iif\ $\cup_{j\in \{1,\dots,n\}\setminus \{i\}}S_{v_j}=\cU$ \tthen
\\
& & {\bf 3.2.1}
\hspace*{0.05in}
run the algorithm of Johnson in~\cite{J74} for this instance of 
\\
& &
	 \hspace*{0.8in}
		set-cover
	  giving a
		solution $\cI\subseteq \{1,\dots,n\}\setminus\{i\}$
\\
& & {\bf 3.2.2}
\hspace*{0.05in}
        $V'=\left\{ \, v_j \,|\, j\in\cI \, \right\}$
\\
& & {\bf 3.2.3}
\hspace*{0.05in}
   \iif\
   $\big( \, |V'|<\hatlopteqone \, \big)$
   \tthen\
   $\,\,\,\hatlopteqone\leftarrow \left|V'\right|$ ;
   $\hatvopteqone\leftarrow V'$
\\
{\bf 4.} &
\multicolumn{2}{l}{
\rreturn\ $\hatlopteqone$ and $\hatvopteqone$ as our solution
}
\end{longtable}
\smallskip

\subsection{Run-time analyses and implementations of Algorithms~I~and~II}

Both Algorithm~I and Algorithm~II use the all-pairs-shortest-path (\apsp) computation, 
and this is the step that dominates the theoretical worst-case
running time of both the algorithms. The following algorithmic approaches are possible 
for the all-pairs-shortest-path step:
\begin{itemize}
\item
For the classical Floyd-Warshall algorithm for \apsp~\cite{CLR90},
the theoretical worst-case running time of 
is $O(n^3)$ when $n$ is the number of nodes in the network.
In practice, for larger networks the running time of the 
Floyd-Warshall algorithm for \apsp
can often be improved by using algorithmic engineering tricks such as 
early termination criteria that are known 
in the algorithms community.

For our networks, we found the Floyd-Warshall algorithm with appropriate data structures and 
algorithmic engineering techniques to be sufficient; one reason for this could be 
that most of our networks, like many other real-world networks, have a small diameter and thus
some computational steps in the Floyd-Warshall algorithm can often be skipped (the diameter of 
a network can be computed in worst-case $o(n^3)$ time~\cite{Y11} and in just $O(m)$ time in practice
for many real-world networks~\cite{CGH13}).
\item
Repeatedly running \emph{breadth-first-search}~\cite{CLR90} from each node gives a solution 
of \apsp with a worst-case running time of $O(mn)$, which is better than $O(n^3)$ if $m=o(n^2)$, \IE, 
the network is sparse.
\item
For specific types of networks,
practitioners also consider using other algorithmic approaches, such as 
repeated use of Dijkstra's single-source shortest path or Johnson's algorithm~\cite{CLR90},
if they are run faster.
Both these algorithms have a worst-case running time of 
$O(n^2\log n+nm)$ where $m$ is the number of edges, and therefore run faster than 
Floyd-Warshall algorithm in the worst case if $m=o(n^2)$. 
\item
Using graph compression techniques, it is possible to design a 
$O(n^3/\log n)$ worst-case time algorithm for \apsp~\cite{FM95}.
\item
Using fast matrix multiplication algorithms, 
\apsp can be solved in $O(n^{2.376})$ time~\cite{GM97,GM97-2,S95}
using Coppersmith and Winograd's matrix multiplication result~\cite{CW90}.
\end{itemize}
For increasing the efficiency and speed of the algorithms we used various data structures 
such as \emph{STL nested maps} and \emph{vectors} to improve comparisons and lookup operations. 
Furthermore, for Algorithm~I, 
we prematurely terminate the algorithm 
if $|\vopt|$ reaches $1$
as $1$ is the smallest value of the size of attacker nodes.    

Finally, just like the measures in this article, the \apsp computation is \emph{unavoidable} for 
a large variety of other geodesic-based 
network properties that are often used for real networks such as the \emph{betweenness centrality},
\emph{closeness centrality}
or \emph{Gromov-hyperbolicity} measure,
and there is a vast amount of literature that apply such measures to large networks
(\EG, see~\cite{BG14,WF94,ADM14,N03,N01,HKYH02}).

\subsection{Scalability of the privacy measure with respect to the size of network}

We have tested computation of the privacy measures for graphs up to $1000$ nodes.
For Algorithm~I, we found that the running time for computing the measure 
for an individual network ranges from $1$ minute or less (for smaller sparser networks) 
to about $10$ to $20$ minutes (for larger denser networks). 
For Algorithm-II the running time was mostly in the order of a few minutes. 

However, for much larger networks than what has been used in this paper, we would
recommend a more careful implementation, specially for Algorithm~I, 
to achieve a more time efficient implementation. 
Towards this goal, we provide the following suggestions in relation to
computing the measures for larger networks:
\begin{itemize}
\item
For larger networks, 
it would be advisable to use the fastest possible implementation of the all-pairs-shortest-paths
algorithm. This is a well-known problem that admits a variety of algorithms some of which are especially
more efficient on non-dense networks and moreover
in practice the running times of many of these algorithms 
can be significantly improved by using several algorithmic engineering tricks (early termination criteria, 
efficient data structures etc.) that are known in the algorithmic implementation 
community.
Also, if the same network is used for more than one privacy measure
computation, it is certainly advisable to store the all-pairs-shortest-path data and re-use them instead
of computing them afresh every time. 
\item
Although our simulation did not need it, for larger networks the relevant set operations needed in 
Algorithms~I~and~II 
can be implemented more efficiently, for example using the well-known data structures for disjoint sets
(\EG, see~\cite{GI91} for a survey).
\item
For extremely large networks, say dense networks containing millions of nodes,
it may be advisable to use a suitable sampling method such as in~\cite{LF06} to sample appropriate sub-graphs
of smaller size, and use the measures computed on these sub-graphs to 
statistically estimate the value of the measures on the entire graph.
\end{itemize}

\subsubsection{Synthetic networks: models and algorithmic generations}

\emph{Unfortunately, there is \emph{no} single universally agreed upon synthetic network model that 
faithfully reproduces all networks in various application domains (\EG, see~\cite{SWM05,KW06,AYS16}).
In fact, there are some results that cast doubt if a true generative network model can even be known unambiguously}.
Thus, it is very customary in the network research community 
to draw conclusions of the following type:

\begin{quote}
``For those real-world networks generated by such-and-such model, we can conclude that $\ldots\ldots$''
\end{quote}

\noindent
We use two major types of synthetic networks, namely the 
\emph{Erd\"{o}s-R\'{e}nyi random networks}
and 
the \emph{scale-free random networks} generated by the 
Bar\'{a}basi-Albert {\em preferential-attachment} model~\cite{BaAl99}.  
Although the 
Erd\"{o}s-R\'{e}nyi network model has been used by prior network researchers
as a real-network model in several application domains (\EG, see~\cite{A10,GK08,MGS09,CNSW00})
it is also known that this particular model is probably not very good a model 
for real networks in many other application domains.
Thus, we also consider networks generated by the 
scale-free random network model 
which is more widely considered to be a real-network model in many network applications (\EG, see
~\cite{BaAl99,ACM11,CMS10,W02,Albert-Barabasi-2002}).

\medskip
\noindent
\textbf{Erd\"{o}s-R\'{e}nyi model}
This is 
the classical undirected
Erd\"{o}s-R\'{e}nyi model $G(n,p)$, where $n$ is the number of nodes and every possible edge in the network
is selected independently with a probability of $p$.
The average degree of any node in $G(n,p)$ is $(n-1)p\approx np$, leading to $\frac{n(n-1)p}{2}\approx\frac{n^2p}{2}$ 
as the average number of edges
in the network.
Our privacy measures assume that the given graph is connected since one connected component has no influence 
on the privacy of another connected component. Thus, it is imperative to select only those combinations of $n$ and $p$
that keeps the graph connected by keeping the average degree of every node to be at least $1$. 
However, 
we actually need to make sure that the average degree is \emph{at least} $2$
since, for example, $\lopteqone$ is trivially equal to $1$ otherwise.
This implies that at the very least we must ensure that $(n-1)p\geq 2$, or \emph{roughly} $np\geq 2$.
However, in practice, while generating the actual random networks one may need to 
select a $p$ that is slightly higher (\textbf{in our case, $np\geq 2.5$}). Note that the giant-component formation in
ER networks happens around $np\approx 1$, so we are indeed further away from this phenomenon 
where slight variations in $p$ cause abrupt changes in topological behavior of the network.
We used the following four combinations of $n$ and $p$ to generate our synthetic networks
to capture a smaller average degree of $2.5$, a modest average degree of $5$
and a larger average degree of $10$:

\begin{center}
\setlength{\tabcolsep}{5pt}
\begin{tabular}{l}
\begin{tabular}{r c l} $n$ & $=$ & $500$ \\ $p$ & $=$ & $0.005$ \\ $np$ & $=$ & $2.5$ \end{tabular}
$\,\,\,\,\,\,$
\begin{tabular}{r c l} $n$ & $=$ & $500$ \\ $p$ & $=$ & $0.01$ \\ $np$ & $=$ & $5$ \end{tabular}
$\,\,\,\,\,\,$
\begin{tabular}{r c l} $n$ & $=$ & $1000$ \\ $p$ & $=$ & $0.005$ \\ $np$ & $=$ & $5$ \end{tabular}
$\,\,\,\,\,\,$
\begin{tabular}{r c l} $n$ & $=$ & $1000$ \\ $p$ & $=$ & $0.01$ \\ $np$ & $=$ & $10$ \end{tabular}
\end{tabular}
\setlength{\tabcolsep}{6pt}
\end{center}

For $n=500$ (respectively, for $n=1000$)
we generated $1000$ random networks (respectively, $100$ random 
networks) for each corresponding value of $p$, and then calculated relevant statistics using Algorithms~I~and~II.

\medskip
\noindent
\textbf{Scale-free model}
We use the Bar\'{a}basi-Albert {\em preferential-attachment} model
~\cite{BaAl99} 
to generate random scale-free
networks. The algorithm for generating a random scale-free 
$G(n,q)$,where $n$ is number of nodes and $q\ll n$ is the number of connections each new node makes, is as follows:
\begin{itemize}
\item  
Initialize $G$ to have $q$ nodes and \emph{no} edges.
Add these nodes to a ``{\em list of repeated nodes}''.

\item
Repeat the following steps till $G$ has $n$ nodes: 
\begin{itemize}
\item
Randomly select $q$ distinct nodes, say $u_1,\dots,u_q$, from the \emph{list of repeated nodes}. 

\item
Add a new node $w$ and undirected edges $\{w,u_1\},\dots,\{w,u_q\}$ in $G$.

\item
Add $w$ and $u_1,\dots,u_q$ to the current \emph{list of repeated nodes}.
\end{itemize}
\end{itemize}
The larger the $q$ is, the more dense is the network $G(n,q)$.
We used the following four combinations of $n$ and $q$ to generate our synthetic scale-free networks: 

\begin{center}
\begin{tabular}{l}
\begin{tabular}{r c l} $n$ & $=$ & $500$ \\ $q$ & $=$ & $5$ \end{tabular}
$\,\,\,\,\,\,$
\begin{tabular}{r c l} $n$ & $=$ & $500$ \\ $q$ & $=$ & $10$ \end{tabular}
$\,\,\,\,\,\,$
\begin{tabular}{r c l} $n$ & $=$ & $1000$ \\ $q$ & $=$ & $5$ \end{tabular}
$\,\,\,\,\,\,$
\begin{tabular}{r c l} $n$ & $=$ & $1000$ \\ $q$ & $=$ & $10$ \end{tabular}
\end{tabular}
\end{center}

%
For $n=500$ (respectively, for $n=1000$)
we generated $1000$ random networks (respectively, $100$ random 
networks) for each corresponding value of $q$, and then calculated relevant statistics using Algorithms~I~and~II.

\subsubsection{Real networks}
\label{sec-real-networks}

\smallskip
\begin{table}[h]
\begin{adjustwidth}{0in}{0in} 
\centering
\caption{
List of real social networks studied in this paper.}
\begin{tabular}{p{0.345\textwidth} r r p{0.46\textwidth} }
\toprule
\multicolumn{1}{c}{{\bf Name}} &  \multicolumn{2}{c}{{\bf \# of}} & \multicolumn{1}{c}{\bf Description}
\\
& \multicolumn{1}{c}{\textbf{nodes}} & \multicolumn{1}{c}{\textbf{edges}} &
\\
\midrule
{\bf (A)}
Zachary Karate Club~\cite{A} & $34$ & $78$ &
Network of friendships between $34$ members
of a karate club at a US university in the $1970$s
\\
\midrule
{\bf (B)}
San Juan Community~\cite{B} & $75$ & $144$ &
Network for visiting relations between
families living in farms in the
neighborhood San Juan Sur, Costa Rica, $1948$
\\
\midrule
{\bf (C)}
Jazz Musician Network~\cite{C} & $198$ & $2842$ &
A social network of Jazz musicians
\\
\midrule
{\bf (D)}
University Rovira i Virgili emails~\cite{D} & $1133$ & $10903$ &
the network of e-mail interchanges
between members of the University
Rovira i Virgili
\\
\midrule
{\bf (E)}
Enron Email Data set~\cite{E} & $1088$ & $1767$ &
Enron email network
\\
\midrule
{\bf (F)}
Email Eu core~\cite{F} & $986$ & $24989$ &
Emails from a large European research institution
\\
\midrule
{\bf (G)}
UC Irvine College Message platform~\cite{G} & $1896$ & $59835$ &
Messages on a Facebook-like platform at UC-Irvine
\\
\midrule
%
%
\hl
{
{\bf (H)}
Hamsterster friendships
  } 
~\cite{HHH} 
	& \hl{$1788$} & \hl{$12476$} &
\hl
{
This Network contains friendships between users of
the website \url{hamsterster.com}
}
\\
\bottomrule
\end{tabular}
\label{L1}
\end{adjustwidth}
\end{table}

Table~\ref{L1} shows the list of eight well-known unweighted social networks that we investigated.
All the networks except one
were undirected;
for the only directed
{\em UC Irvine College Message platform}
network, we ignored the direction of edges.
For each network the \emph{largest} connected component was selected and tested.

\subsubsection{Results for real networks in Table~\ref{L1}}

\medskip
\noindent
\textbf{Results for \nokmad\ and \mad}
Table~\ref{L2} shows 
the results for \nokmad\ via applying Algorithm~I to these networks. 
From these results we may conclude:

\begin{quote}
\begin{description}
\item[{\LARGE\ding{172}}]
For all networks \emph{except}
the ``Enron Email Data'' network, 
an attacker needs to control \emph{only one} suitable node of the network to 
uniquely re-identify
(based on the metric representation) 
a significant percentage of nodes in the network 
(ranging from 
$2.6\%$ of nodes for the 
``University Rovira i Virgili emails'' network 
to $26.5\%$ of nodes for the 
``Zachary Karate Club'' network). 
\item[{\LARGE\ding{173}}]
For all networks \emph{except}
the ``Enron Email Data'' network, 
the minimum privacy violation probability guarantee is significantly further from zero 
(ranging from 
$0.019$ for the 
``UC Irvine College Message platform''
network 
to 
$0.25$ for the 
``Hamsterster friendships'' 
network).
The minimum privacy violation probability guarantee 
for the 
``Hamsterster friendships'' 
network
is significantly higher than all other networks.
\item[{\LARGE\ding{174}}]
The 
``Zachary Karate Club''
and the 
``San Juan Community'' networks 
are \emph{more} vulnerable to privacy attacks in terms of the percentage of nodes in the networks whose privacy
can be violated by the adversary.
\end{description}
\end{quote}

\begin{table}[!ht]
\begin{adjustwidth}{0in}{0in} 
\centering
\caption{Results for \nokmad\ using Algorithm~I. 
$n$ is the number of nodes and 
$\kopt$ is the largest value of $k$
such that 
$\voptgt\neq\emptyset$ (\emph{cf}.\ Problem~\ref{prob1}).
}
\begin{tabular}{l r r c r r }
\toprule
\multicolumn{1}{c}{\bf Name} &
   \multicolumn{1}{c}{$n$} & 
	 $\kopt$ & $\popt=\nicefrac{1}{\kopt}$ & 
	 \multicolumn{1}{c}{$\mathpzc{L}_{\mathrm{opt}}^{\geq \kopt}=\mathpzc{L}_{\mathrm{opt}}^{=\kopt}$}
	 & 
	 \multicolumn{1}{c}{$\frac{\kopt}{n}$}
\\
\midrule
{\bf (A)} 
Zachary Karate Club &
  $34$ & 
	$9\,\,$ & $0.111$ & $1$
	 \hspace*{0.3in}
	& $26.5\%\,\,\,\,\,$
\\
\midrule
{\bf (B)} 
San Juan Community &
  $75$ & 
	$7\,\,$ & $0.143$ & $1$
	 \hspace*{0.3in}
	& $9.3\%\,\,\,\,\,$
\\
\midrule
{\bf (C)} 
Jazz Musician Network &
  $198$ & 
	$12\,\,$ & $0.084$ & $1$
	 \hspace*{0.3in}
	& $6.0\%\,\,\,\,\,$
\\
\midrule
{\bf (D)} 
University Rovira i Virgili emails &
  $1133$ & 
	$29\,\,$ & $0.035$ & $1$
	 \hspace*{0.3in}
	& $2.6\%\,\,\,\,\,$
\\
\midrule
{\bf (E)} 
Enron Email Data set &
  $1088$ & 
	$153\,\,$ & $0.007$ & $935$
	 \hspace*{0.3in}
	& $14.1\%\,\,\,\,\,$
\\
\midrule
{\bf (F)} 
Email Eu core &
  $986$ & 
	$39\,\,$ & $0.026$ & $1$
	 \hspace*{0.3in}
	& $3.4\%\,\,\,\,\,$
\\
\midrule
{\bf (G)} 
UC Irvine College Message platform &
  $1896$ & 
	$55\,\,$ & $0.019$ & $1$
	 \hspace*{0.3in}
	& $2.9\%\,\,\,\,\,$
\\
\midrule
\hl{{\bf (H)} Hamsterster friendships} &
  \hl{$1788$} & 
	\hl{$4\,\,$} & \hl{$0.25$} & \hl{$1$}
	 \hspace*{0.3in}
	& \hl{$0.22\%\,\,\,\,\,$}
\\
\bottomrule
\end{tabular}
\label{L2}
\end{adjustwidth}
\end{table}

For the ``Enron Email Data'' network,
$\mathpzc{L}_{\mathrm{opt}}^{\geq \kopt}=935$ implies that 
even to achieve a modest value of $\popt=0.007$
an adversary needs to control 
a large percentage (at least $\frac{935\times 100}{1088}\%\approx 86\%$) of its nodes, 
a possibility unlikely to happen in practice. 
Thus, we continue further investigation about this network to check if a value of $k$ \emph{somewhat} smaller than $\kopt$ 
may allow a \emph{sufficiently steep} decline in the number of nodes that the attacker need to control, 
and report 
the values of $\loptgt$ corresponding to relevant values of $k>1$
in Table~\ref{L3}.
As can be seen, 
the values of 
$\loptgt$
does not decline unless $k$ is really further away from $\kopt$, leading us to conclude the following: 

\begin{quote}
\begin{description}
\item[{\LARGE\ding{175}}]
For 
the ``Enron Email Data'' network,
privacy violation of a large number of nodes of the network 
by an attacker cannot be guaranteed 
in a \emph{practical} sense (\IE, without gaining control of a large number of nodes).
\end{description}
\end{quote}

%
\setlength{\tabcolsep}{5pt}
\begin{table}[!ht]
\begin{adjustwidth}{0in}{0in} 
\centering
\caption{
Values of $\loptgt$ corresponding to values for $k>1$ for ``Enron Email Data'' network.
Only those values of $k>1$ for which $\loptgt\neq \mathpzc{L}_{\mathrm{opt}}^{\geq k-1}$ are shown.}
\begin{tabular}{l l}
\toprule
\begin{tabular}{l l }
{\bf (E)} 
Enron Email Data set &
\begin{tabular}{r c c c c c c c c c }
$k$ &  $4$ & $5$ & $10$ & $20$ & $40$ & $60$ & $100$ & $120$  & $153$
\\
\midrule
$p_k=\nicefrac{1}{k}$ &  $0.25$ & $0.2$ & $0.1$ & $0.05$ & $0.025$ & $0.017$ & $0.01$ & $0.009$  & $0.007$
\\
\midrule
$\loptgt$ & $1$ & $334$ &  $463$ & $567$ & $683$ & $842$ & $935$ & $935$ & $935$
\end{tabular}
\end{tabular}
\\
\bottomrule
\end{tabular}
\label{L3}
\end{adjustwidth}
\end{table}
\setlength{\tabcolsep}{6pt}

\medskip
\noindent
\textbf{Results for \eqmadone}
Algorithm~II returns
$\lopteqone=1$ for all of our networks except the 
``Hamsterster friendships'' network.
For the 
``Hamsterster friendships'' network, 
Algorithm~II returns
$\lopteqone=2$.
Thus, we conclude:

\begin{quote}
\begin{description}
\item[{\LARGE\ding{176}}]
For all the real networks except the 
``Hamsterster friendships'' network, an adversary controlling \emph{just one} suitable node
may uniquely re-identify 
(based on the metric representation) 
one other node in the network with certainty (\IE, with a probability of $1$).
For the 
``Hamsterster friendships'' network, 
the same conclusion holds provided the adversary 
controls two suitable nodes.
\end{description}
\end{quote}

\subsubsection{Results for Erd\"{o}s-R\'{e}nyi synthetic networks}

\medskip
\noindent
\textbf{Results for \mad}
Table~\ref{L2-2} shows 
the results for \mad\ via applying Algorithm~I to these networks. 
From these results we may conclude:

\begin{quote}
\begin{description}
\item[{\LARGE\ding{177}}]
For \emph{most} synthetic Erd\"{o}s-R\'{e}nyi networks,  
$\kopt$ is a value that is \emph{much smaller} compared to the number of nodes $n$.
Thus, for our synthetic Erd\"{o}s-R\'{e}nyi networks, with high probability
privacy violation of a large number of nodes of the network 
by an attacker \emph{cannot} be achieved. 
\item[{\LARGE\ding{178}}]
The values of $\frac{\kopt}{n}$ for denser Erd\"{o}s-R\'{e}nyi networks (corresponding to $p=0.01$) is about $75\%$ higher 
that those for 
sparser Erd\"{o}s-R\'{e}nyi networks (corresponding to $p=0.005$)
irrespective of the number of nodes. Thus, we conclude that 
our sparser synthetic Erd\"{o}s-R\'{e}nyi networks are more privacy-secure compared to their denser counter-parts. 
\end{description}
\end{quote}

\setlength{\tabcolsep}{4pt}
\begin{table}[!ht]
\begin{adjustwidth}{0in}{0in} 
\centering
\caption{
Results for \mad\ using Algorithm~I for classical Erd\"{o}s-R\'{e}nyi model $G(n,p)$.
$\kopt$ is the largest value of $k$
such that $\voptgt\neq\emptyset$ (cf.\ Problem~\ref{prob1}).
The \%-values indicate the percentage of the generated 
networks for those particular values of $\kopt$
(\EG, for $n=500$ and $p=0.005$, $980$ out of the $1000$ networks have $\kopt\geq 5$).
}
\begin{tabular}{c c  r c c c c c c c c }
\toprule
\multicolumn{2}{c}{Network} &
\\
\multicolumn{2}{c}{parameters} &
\\
\cmidrule{1-2}
$n$ & \hspace*{-0.4in} $p$
&
\\
\midrule
\midrule
\multirow{4}{*}{$500$} & \multirow{4}{*}{\hspace*{-0.1in} $0.005$} & 
         \hspace*{-0.2in}
           $\kopt$ & $\geq 4$ & $\geq 5$ & $\geq 6$ & $\geq 7$ & $\geq 8$ & $\geq 9$ & $\geq 10$ & $>10$ 
\\
\cmidrule{3-11}
 &  & \hspace*{-0.2in}
      $\popt=\nicefrac{1}{\kopt}$ 
        & $\leq 0.25$ & $\leq 0.2$ & $\leq 0.166$ & $\leq 0.142$ & $\leq 0.125$ & $\leq 0.111$ & $\leq 0.1$ & $<0.1$ 
\\
\cmidrule{3-11}
 &  & \hspace*{-0.2in}
      \% of networks 
        & $100\%$ & $98\%$ & $81.8\%$ & $54.6\%$ & $21.5\%$ & $8\%$ & $3\%$ & $1\%$
\\
\cmidrule{3-11}
 &  & 
     \multicolumn{9}{c}{At least $90\%$ of networks have $\kopt\leq 8$ and $\frac{\kopt}{n}\leq 0.016$}
\\ 
\midrule
\midrule
\multirow{4}{*}{$500$} & \multirow{4}{*}{\hspace*{-0.1in} $0.010$} & 
         \hspace*{-0.2in}
     $\kopt$ & $\geq 9$ & $\geq 10$ & $\geq 11$ & $\geq 12$ & $\geq 13$ & $\geq 14$ & $\geq 15$ & $>15$ 
\\
\cmidrule{3-11}
 &  & \hspace*{-0.2in}
  $\popt=\nicefrac{1}{\kopt}$ 
     & $\leq 0.11$ & $\leq 0.1$ & $\leq 0.09$ & $\leq 0.083$ & $\leq 0.077$ & $\leq 0.071$ & $\leq 0.066$ & $<0.066$ 
\\
\cmidrule{3-11}
 &  & \hspace*{-0.2in}
     \% of networks 
 & $100\%$ & $98\%$ & $94\%$ & $81.4\%$ & $49.4\%$ & $21.4\%$ & $6.8\%$ & $0.6\%$
\\
\cmidrule{3-11}
 &  & 
       \multicolumn{9}{c}{At least $90\%$ of networks have $\kopt\leq 14$ and $\frac{\kopt}{n}\leq 0.028$}
\\ 
\midrule
\midrule
\multirow{4}{*}{$1000$} & \multirow{4}{*}{\hspace*{-0.1in} $0.005$} & 
         \hspace*{-0.2in}
      $\kopt$ & $\geq 10$ & $\geq 11$ & $\geq 12$ & $\geq 13$ & $\geq 14$ & $>14$ 
\\
\cmidrule{3-11}
 &  & \hspace*{-0.2in}
  $\popt=\nicefrac{1}{\kopt}$ 
& $\leq 0.1$ & $\leq 0.09$ & $\leq 0.083$ & $\leq 0.077$ & $\leq 0.071$ & $<0.066$ 
\\
\cmidrule{3-11}
 &  & \hspace*{-0.2in}
\% of networks 
& $100\%$ & $99\%$ & $65\%$ & $16\%$ & $7\%$ & $1\%$ 
\\
\cmidrule{3-11}
 &  & 
       \multicolumn{9}{c}{At least $90\%$ of networks have $\kopt\leq 13$ and $\frac{\kopt}{n}\leq 0.013$}
\\
\midrule
\midrule
\multirow{4}{*}{$1000$} & \multirow{4}{*}{\hspace*{-0.1in} $0.010$} & 
         \hspace*{-0.2in}
     $\kopt$ & $\geq 18$ & $\geq 19$ & $\geq 20$ & $\geq 21$ & $\geq 22$ & $\geq 23$ & $\geq 24$ & $>24$ 
\\
\cmidrule{3-11}
 &  & \hspace*{-0.2in}
  $\popt=\nicefrac{1}{\kopt}$ 
& $\leq 0.055$ & $\leq 0.052$ & $\leq 0.05$ & $\leq 0.047$ & $\leq 0.045$ & $\leq 0.043$& $\leq 0.041$ & $<0.041$ 
\\
\cmidrule{3-11}
 &  & \hspace*{-0.2in}
\% of networks 
& $100\%$ & $99\%$ & $90\%$ & $75\%$ & $47\%$ & $26\%$ & $9\%$ & $1\%$ 
\\
\cmidrule{3-11}
 &  & 
       \multicolumn{9}{c}{At least $90\%$ of networks have $\kopt\leq 23$ and $\frac{\kopt}{n}\leq 0.023$}
\\ 
\bottomrule
\end{tabular}
\label{L2-2}
\end{adjustwidth}
\end{table}
\setlength{\tabcolsep}{6pt}

\medskip
\noindent
\textbf{Results for \eqmadone}
Table~\ref{L4} shows the result of our experiments of 
computation of 
$\lopteqone$
using Algorithm~II.
From these results, we conclude:

\begin{quote}
\begin{description}
\item[{\LARGE\ding{179}}]
For our synthetic Erd\"{o}s-R\'{e}nyi networks, with high probability
an adversary controlling \emph{at most two} nodes
may uniquely re-identify 
(based on the metric representation) 
\emph{at least} one other node in the network.
\end{description}
\end{quote}

\begin{table}[!ht]
\centering
\caption{
Results for \eqmadone\ using Algorithm~\rom{2} for classical
Erd\"{o}s-R\'{e}nyi model $G(n,p)$.
The \%-values indicate the percentage of the generated networks that have the corresponding value
of $\lopteqone$ (\EG, for $n=500$ and $p=0.01$, $920$ out of the $1000$ networks have $\lopteqone=1$).}
\begin{tabular}{c c || r r r}
\toprule
\multicolumn{2}{c||}{Network parameters} & \multicolumn{3}{c}{$\lopteqone$}
\\
\midrule
$n$ & $p$ 
&
\multicolumn{1}{c}{$1$} & \multicolumn{1}{c}{$2$} & \multicolumn{1}{c}{$>2$}
\\
\midrule
$500$ & $0.01$ &  $92$\% &    $7$\%&  $1$\%
\\
\midrule
$500$ & $0.005$ &  $5.9$\% &    $89.3$\%&  $4.8$\%
\\
\midrule
$1000$ & $0.01$ & $8$\% & $90$\%   & $2$\%
\\
\midrule
$1000$ & $0.005$ & $5$\% & $93$\%   & $1$\%
\\
\bottomrule
\end{tabular}
\label{L4}
\end{table}

%
\setlength{\tabcolsep}{3pt}
\begin{table}[hbt]
\begin{adjustwidth}{0in}{0in} 
\centering
\caption{
Results for \mad\ using Algorithm~I for the Bar\'{a}basi-Albert preferential-attachment scale-free model $G(n,q)$.
$\kopt$ is the largest value of $k$
such that $\voptgt\neq\emptyset$ (cf.\ Problem~\ref{prob1}).
The \%-values indicate the percentage of the generated 
networks for those particular values of $\kopt$
(\EG, for $n=500$ and $q=5$, $990$ out of the $1000$ networks have $\kopt\geq 50$).
}
\begin{tabular}{c c   @{\hskip -0.0in}  r    c c c c c c c c }
\toprule
\multicolumn{2}{c}{Network} &
\\
\multicolumn{2}{c}{parameters} &
\\
\cmidrule{1-2}
$n$ & \hspace*{-0.4in} $q$
&
\\
\midrule
\midrule
\multirow{4}{*}{$500$} & \multirow{4}{*}{\hspace*{-0.25in} $5$} & 
         \hspace*{-0.3in}
           $\kopt$ & $\geq 49$ & $\geq50$ & $\geq55$ & $\geq60$ & $\geq 65$ & $\geq 70$ & $> 70$ &
\\
\cmidrule{3-11}
 &  & \hspace*{-0.3in}
      $\popt=\nicefrac{1}{\kopt}$ 
          & $\leq0.0204$ & $\leq0.02$ & $\leq0.018$ & $\leq0.016$ & $\leq0.015$ & $\leq0.014$ & $< 0.014$ & 
\\
\cmidrule{3-11}
 &  & \hspace*{-0.3in}
      \% of networks 
          & $100\%$ & $99\%$ & $97\%$ & $89\%$ & $42\%$ & $10\%$ &  $6\%$& 
\\
\cmidrule{3-11}
 &  & 
     \multicolumn{9}{c}{At least $90\%$ of networks have $\kopt\leq 65$ and $\frac{\kopt}{n}\leq 0.13$}
\\ 
\midrule
\midrule
\multirow{4}{*}{$500$} & \multirow{4}{*}{\hspace*{-0.25in} $10$} & 
         \hspace*{-0.3in}
       $\kopt$ & $\geq 45$ & $\geq 60$ & $\geq 80$ & $\geq 100$ & $\geq 120$ & $\geq 140$ & $> 140$   & 
\\
\cmidrule{3-11}
 &  & \hspace*{-0.3in}
  $\popt=\nicefrac{1}{\kopt}$ 
     & $\leq 0.022$ & $\leq 0.016$ & $\leq 0.0125$ & $\leq 0.001$ & $\leq 0.008$ & $\leq 0.007$ & $< 0.007$ & 
\\
\cmidrule{3-11}
 &  & \hspace*{-0.3in}
     \% of networks 
   & $100\%$ & $50\%$ & $48\%$ & $47\%$   & $27\%$   & $5\%$    & $4\%$   & 
\\
\cmidrule{3-11}
 &  & 
       \multicolumn{9}{c}{At least $95\%$ of networks have $\kopt\leq 120$ and $\frac{\kopt}{n}\leq 0.24$}
\\ 
\midrule
\midrule
\multirow{4}{*}{$1000$} & \multirow{4}{*}{\hspace*{-0.25in} $5$} & 
         \hspace*{-0.3in}
      $\kopt$ & $\geq 88$ & $\geq 90$ & $\geq 100$ & $\geq 110$ & $\geq 120$ & $\geq 130$  & $\geq 135$
\\
\cmidrule{3-11}
 &  & \hspace*{-0.3in}
  $\popt=\nicefrac{1}{\kopt}$ 
     & $\leq 0.011$ & $\leq 0.010$ & $\leq 0.001$ & $\leq 0.009$ & $\leq 0.008$ &$\leq 0.007$ & $\leq 0.0074$ 
\\
\cmidrule{3-11}
 &  & \hspace*{-0.3in}
\% of networks 
        & $100\%$ & $98\%$ & $94\%$ & $66\%$ & $32\%$ & $11\%$ & $1\%$
\\
\cmidrule{3-11}
 &  & 
       \multicolumn{9}{c}{At least $89\%$ of networks have $\kopt\leq 120$ and $\frac{\kopt}{n}\leq 0.12$}
\\
\midrule
\midrule
\multirow{4}{*}{$1000$} & \multirow{4}{*}{\hspace*{-0.25in} $10$} & 
         \hspace*{-0.3in}
     $\kopt$ & $\geq 86$ & $\geq 88$ & $\geq 90$ & $\geq 92$ & $\geq 94$ & $\geq 96$ & $\geq 98 $ & $>100$  
\\
\cmidrule{3-11}
 &  & \hspace*{-0.3in}
  $\popt=\nicefrac{1}{\kopt}$ 
      & $\leq 0.0116$ & $\leq 0.0113$ & $\leq 0.0111$ & $\leq 0.0108$ & $\leq 0.0106$ & $\leq 0.0104$ & $\leq 0.0102$ & $<0.001$ 
\\
\cmidrule{3-11}
 &  & \hspace*{-0.3in}
\% of networks 
         & $100\%$ & $77\%$ & $67\%$ & $56\%$ & $43\%$ & $30\%$ & $13\%$ & $3\%$ 
\\
\cmidrule{3-11}
 &  & 
       \multicolumn{9}{c}{At least $87\%$ of networks have $\kopt\leq 96$ and $\frac{\kopt}{n}\leq 0.096$}
\\ 
\bottomrule
\end{tabular}
%
\label{L5}
\end{adjustwidth}
\end{table}
\setlength{\tabcolsep}{6pt}

\subsubsection{Results for scale-free synthetic networks}

\medskip
\noindent
\textbf{Results for \mad}
Table~\ref{L5} shows 
the results for \mad\ via applying Algorithm~I to these networks. 
From these results we may conclude:

\begin{quote}
\begin{description}
\item[{\LARGE\ding{180}}]
The value of $\kopt$ relative to the size $n$ of the network 
is much larger for synthetic scale-free networks compared to those for the synthetic 
Erd\"{o}s-R\'{e}nyi networks.
Thus, compared to synthetic Erd\"{o}s-R\'{e}nyi networks,
synthetic scale-free networks
may allow 
privacy violation of a larger number of nodes of the network 
by an attacker.
\item[{\LARGE\ding{181}}]
Unlike the synthetic Erd\"{o}s-R\'{e}nyi networks, 
the values of $\frac{\kopt}{n}$ for denser scale-free networks (corresponding to $q=10$) may be smaller or larger than 
those for 
sparser scale-free networks (corresponding to $q=5$).
Thus, density of scale-free networks does not seem to be well-correlated to 
privacy-security of these networks.
\end{description}
\end{quote}

\medskip
\noindent
\textbf{Results for \eqmadone}
Table~\ref{L6} shows the result of our experiments of 
computation of 
$\lopteqone$
using Algorithm~II.
From these results, we conclude:

\begin{quote}
\begin{description}
\item[\circled{$\pmb{11}$}]
Similar to synthetic synthetic Erd\"{o}s-R\'{e}nyi networks, 
for synthetic scale-free networks also 
with high probability
an adversary controlling \emph{at most two} nodes
may uniquely re-identify 
(based on the metric representation) 
\emph{at least} one other node in the network.
\end{description}
\end{quote}

\begin{table}[!ht]
\centering
\caption{
Results for \eqmadone\ using Algorithm~\rom{2} for 
the Bar\'{a}basi-Albert preferential-attachment scale-free model $G(n,q)$.
The \%-values indicate the percentage of the generated networks that have the corresponding value
of $\lopteqone$ (\EG, for $n=500$ and $q=5$, $990$ out of the $1000$ networks have $\lopteqone=2$).
}
\begin{tabular}{c c || r r}
\toprule
\multicolumn{2}{c||}{Network parameters} & \multicolumn{2}{c}{$\lopteqone$}
\\
\midrule
$n$ & $q$ 
&
\multicolumn{1}{c}{$2$} & \multicolumn{1}{c}{$>2$}
\\
\midrule
$500$ & $5$ &  $99$\% &  $1$\%
\\
\midrule
$500$ & $10$ &  $99.5$\% &  $0.5$\%
\\
\midrule
$1000$ & $5$ & $99$\%   & $1$\%
\\
\midrule
$1000$ & $10$ & $99$\%   & $1$\%
\\
\bottomrule
\end{tabular}
\label{L6}
\end{table}

\section{Conclusion}

Rapid evolution of popular social networks such as Facebook and Twitter have rendered modern society 
heavily dependent on such virtual platforms for their day-to-day operation. 
However, the many benefits accrued by such online networked systems are not necessarily cost-free 
since a malicious entity may compromise privacy of users of these social networks for harmful purposes 
that may result in the disclosure of sensitive attributes of these networks. In this article, we investigated, both 
theoretically and empirically, quantifications of privacy violation measures of large networks under active attacks. 
Our theoretical result indicates that the network manager responsible for prevention of privacy violation must be very 
careful in designing the network if its topology does not contain a cycle, while our empirical results 
shed light on privacy violation properties of \hl{eight} real social networks as well as synthetic networks 
generated by the classical Erd\"{o}-R\`{e}nyi model. 
We believe that our results will stimulate much needed further research on quantifying and computing privacy measures for networks.

\section{Acknowledgements}

We thank the anonymous reviewers for their helpful comments.
B.D. and N.M. thankfully acknowledges 
partially support from NSF grant IIS-1160995.
This research was partially done while I.G.Y. was visiting the University of Illinois at Chicago, USA, supported by 
``Ministerio de Educaci\'on, Cultura y Deporte'', Spain, under the 
``Jos\'e Castillejo'' program for young researchers (reference number: CAS15/00007).

\end{document}